\documentclass[aps,prc,twocolumn,floatfix,superscriptaddress,nofootinbib]{revtex4-2}

\usepackage{graphicx}
\usepackage[urlcolor=green]{hyperref}
\usepackage{amsmath,bm}
\usepackage{physics}
\usepackage{multirow}

\usepackage{color}
\usepackage[dvipsnames]{xcolor}

\newcommand{\McDipper}{{\sc McDipper}}
\newcommand{\iSS}{{\sc iSS}}

\newcommand{\MUSIC}{{\sc music}}

\renewcommand{\dd}{{\rm d}}
\newcommand{\ee}{{\rm e}}
\newcommand{\ii}{{\rm i}}
\newcommand{\snn}{s_{_\mathrm{NN}}}
\newcommand{\Nev}{N_{\mathrm{ev}}}

\begin{document}

\author{Renata Krupczak} \email{rkrupczak@physik.uni-bielefeld.de}
\affiliation{Fakult\"at f\"ur Physik, Universit\"at Bielefeld, D-33615 Bielefeld, Germany}
\author{Nicolas Borghini} 
\affiliation{Fakult\"at f\"ur Physik, Universit\"at Bielefeld, D-33615 Bielefeld, Germany}
\author{Hendrik Roch}
\affiliation{Department of Physics, University of Jyv\"askyl\"a, P.O. Box 35, 40014 University of Jyv\"askyl\"a, Finland}
\affiliation{Helsinki Institute of Physics, P.O. Box 64, 00014 University of Helsinki, Finland}

\title{Event-by-event fluctuations of elliptic flow in ultrarelativistic O+O collisions}

\begin{abstract}
    We study O+O collisions at $\sqrt{\snn} = 5.36$~TeV within a fully three-dimensional \McDipper+\MUSIC\ model, which allows us to describe the experimentally measured dependence of charged hadron multiplicity on centrality and pseudorapidity.   
    We show that the initial elliptical eccentricity is mainly driven by the fluctuations of the energy deposition and thereby varies considerably event-by-event within a fixed centrality class.
    This also holds for elliptic flow $v_2$, whose origin in O+O thus differs from that in collisions of heavy nuclei. 
    Using a decomposition of initial states in an average event and uncorrelated modes, we find that despite the large size of fluctuations we can reproduce the joint probability distribution of eccentricity and elliptic flow with a reasonable accuracy with only a small set of fluctuation modes. 
\end{abstract}

\maketitle

\section{Introduction}

In July 2025, oxygen (${}^{16}$O) collisions took place at the CERN Large Hadron Collider (LHC).
One of the goals was to produce with high statistics a system with a ``size'', quantified for instance by the charged hadron multiplicity, intermediate between the large systems created in Pb+Pb or Xe+Xe collisions and the small ones produced in p+p or p+A. 
This newly available system size should then make it possible to investigate the onset of phenomena that are present in large systems but absent in minimum-bias p+p collisions, shedding light on the properties of the medium created in the former case, but also to study collective phenomena, like transverse flow, that have been observed over the whole range of system sizes. 
Indeed, one of the first reported measurement was that of anisotropic flow, namely elliptic ($v_2$), triangular ($v_3$) and quadrangular ($v_4$) flows~\cite{ALICE:2025luc,ATLAS:2025nnt,CMS:2025tga}. 

In collisions of large nuclei at ultrarelativistic energies, the prevalent view is that anisotropic transverse flow is a consequence of an asymmetry in the initial geometry of the system (in the transverse plane), characterized by eccentricities. 
This asymmetry is then converted into an anisotropy in momentum space by the dynamical evolution~\cite{Ollitrault:1992bk,Heinz:2013th}. 
In his pioneering article, Ollitrault considered the (elliptic) eccentricity $\varepsilon_2$ from the almond-shaped overlap region of the two colliding nuclei, viewed as structureless spheres, when the impact parameter of the collision is finite. 
It was much later understood that the event-by-event change in the positions of the nucleons inside the colliding nuclei also contributes to the initial asymmetry~\cite{Alver:2010gr,Luzum:2013yya}. 
This contribution is the main source of triangularity ($\varepsilon_3$) in the initial state of collisions of identical (and not too deformed) nuclei. 
Yet its contribution to $\varepsilon_2$ is not on a par with that due to the finite impact parameter in collisions of large nuclei. 
Accordingly, in such collisions $\varepsilon_2$ changes significantly with impact parameter --- and thus with centrality, which is correlated to the size of the overlap region ---, while the dependence of $\varepsilon_3$ is much weaker~\cite{Beraudo:2018tpr}.

How different models of the oxygen nucleus affect the initial-state geometry of O+O collisions and final-state anisotropic-flow coefficients has been investigated within a wide range of approaches~\cite{Lim:2018huo,Sievert:2019zjr,Rybczynski:2019adt,Summerfield:2021oex,Behera:2021zhi,Ding:2023ibq,YuanyuanWang:2024sgp,Prasad:2024ahm,Zhang:2024vkh,Mehrabpour:2025rzt,MenonKavumpadikkalRadhakrishnan:2025apq,Loizides:2025ule,Constantin:2025ova,Ito:2026hec}.
For the oxygen nuclei, nucleons were either sampled independently from a two- or three-parameter Fermi distribution, or clustered in $\alpha$-particles, or taken from nucleon configurations from ab initio nuclear-structure computations. 
The nucleon distributions served as input in various initial-state models: IP-Jazma, wounded-nucleon Glauber model, T$_{\rm R}$ENTo, or SMASH initial conditions.
Eventually, the subsequent dynamical evolution involved different models and codes, relying either on viscous hydrodynamics or on a transport approach, or on a combination of both. 

These studies mostly focus on the behavior of the average eccentricities or average flow harmonics within a given centrality class.
In the present paper, we go one step further, and investigate the event-by-event fluctuations of eccentricities and flow coefficients within a centrality class. 
From that, we will argue that O+O collisions are not a scaled-down version of what happens with heavy nuclei. 
Instead, event-by-event fluctuations are already the dominant contribution to elliptical eccentricity, and thus to elliptic flow, which makes such collisions closer to p+p than to Pb+Pb in that respect.

Section~\ref{sec:setup} introduces our three-dimensional (3D) numerical setup for simulating O+O collisions. 
Decomposing the initial state energy-density profiles into an average state and fluctuation modes~\cite{Floerchinger:2013tya,Floerchinger:2018pje,Borghini:2022iym}, which to our knowledge is performed for the first time in a non-boost-invariant scenario, we then present our results for the properties of the average initial state (Sec.~\ref{sec:AvgState}). 
We then turn in Sec.~\ref{sec:Modes} to the discussion of the fluctuation modes, including their impact on initial eccentricity and final-state elliptic flow. 
In Sec.~\ref{sec:v2e2}, we explore how a reduced set of modes can reproduce the main features of the statistics of $\varepsilon_2$ and $v_2$ by comparing to a sample of events. 
Eventually, we present our conclusions in Sec.~\ref{sec:conclusions}.

\section{Event generation}
\label{sec:setup}

In this study, the events are modeled with a 3D \McDipper+\MUSIC+\iSS\ setup, which we now describe. 

The starting point consists of nucleon configurations in the ground state of the oxygen nucleus calculated with the Nuclear Lattice Effective Field Theory (NLEFT) method~\cite{Giacalone:2024luz}. 
Two such configurations are chosen, randomly rotated in three dimensions, and their respective centers-of-mass are separated by a random impact parameter value. 
The resulting nucleon distributions serve as input to the \McDipper\ code~\cite{Garcia-Montero:2023gex}, which is based on the $k_{\rm T}$-factorization limit of the Color Glass Condensate effective field theory. 
Within the GBW (Golec-Biernat--Wusthoff) mode of \McDipper, including thickness fluctuations (with width $\sigma=0.637$) and $N_q = 3$ hot spots to model subnucleonic structure~\cite{Garcia-Montero:2025bpn}, we generate 3D energy density profiles $\{\Phi^{(i)}(x,y,\eta_s)\}$, with the $x$-axis along the impact parameter direction while $\eta_s$ denotes spacetime rapidity.
These profiles are discretized on a finite grid, with a spatial extent of 8.4~fm and a resolution of 0.1~fm in the transverse plane, sufficient to capture the fluctuations generated by \McDipper. 
Along the longitudinal direction, a resolution of 0.2 over the range $\eta_s \in [-10, 10]$ is used for the computations within \McDipper, yet for the initial profiles we only keep the energy deposited at $\eta_s=0$, $\pm 1.5$, $\pm 3$, and $\pm 4.5$ for the fluctuation-mode analysis discussed in Sec.~\ref{sec:Modes}.
The only free parameter in \McDipper\ is a factor $K_g$ that accounts for higher-order corrections to gluon production. 
As explained hereafter, this parameter is fixed by comparison to experimental data on charged hadron multiplicity.

The energy-density profiles from \McDipper\ are used as initial states for a full non-boost-invariant hydrodynamic evolution with \MUSIC~\cite{Schenke:2010nt,Schenke:2010rr,Paquet:2015lta}, starting at proper time $\tau_0 = 0.6$~fm, following Ref.~\cite{Garcia-Montero:2025bpn}. 
This evolution, with a constant specific shear viscosity $\eta/s = 0.16$ and a temperature-dependent bulk viscosity $\zeta/s$ given by the parametrization in Ref.~\cite{Denicol:2009am}, proceeds until the fluid cells reach a temperature of $T = 155$ MeV. 
At this stage, the fluid cell is converted into hadrons using the \iSS\ package~\cite{Shen:2014vra}. 
Since we will only consider global observables, whose values are not significantly modified by rescatterings, we do not employ any afterburner, but only perform hadron decays with \iSS. 
For each hydrodynamic event we produce a large number of \iSS\ oversamples, namely $4 \times 10^5$ or $8 \times 10^5$ depending on the event multiplicity.
Computations requiring the full time evolution of the initial-state density profiles are performed using the hybrid framework CRONOS~\cite{hendrik_roch_2025_17503591}.

\begin{figure}[t]
	\includegraphics[width=0.85\linewidth]{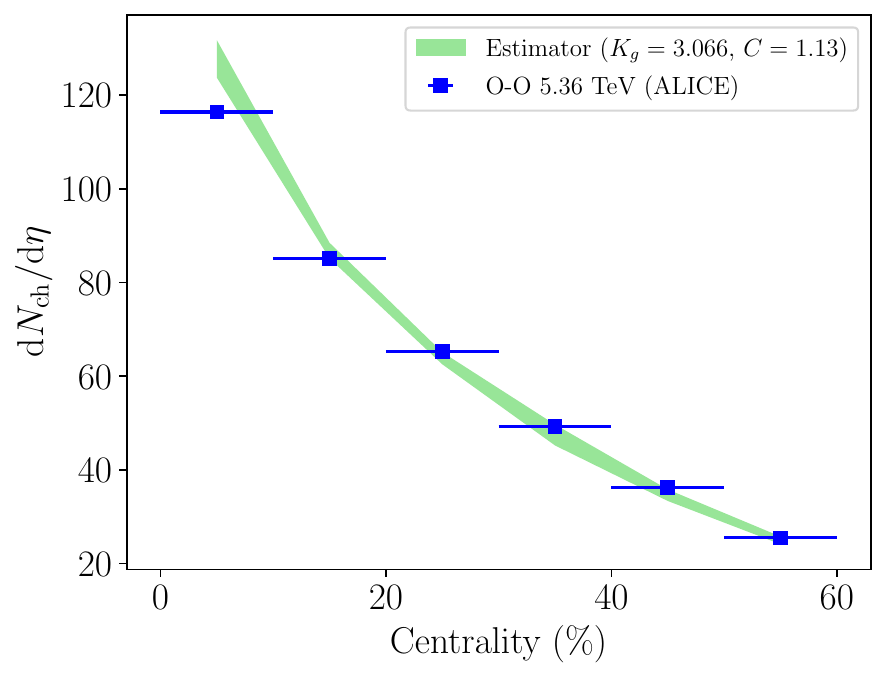}\vspace{-3mm}
	\caption{Centrality dependence of the charged-hadron multiplicity in O+O collisions at $\sqrt{\snn} = 5.36$ TeV. 
		Blue points: preliminary ALICE data~\cite{Modak:2026ymk}. 
		Green curve: result obtained from Eq.~\ref{eq:estimator} with $C = 1.13$ and using $K_g = 3.066$.}
	\label{fig:calibration}
\end{figure}

In the final state of \iSS, one can in particular obtain the charged hadron multiplicity at mid-(pseudo)rapidity, ${\rm d}N_{\rm ch}/{\rm d}\eta$. 
Demanding agreement with the preliminary experimental data from the ALICE Collaboration~\cite{Modak:2026ymk} will fix the only remaining parameter in our approach, namely the $K_g$-parameter in \McDipper.
For that, we need to determine centrality classes in our model. 
Since running 3D hydrodynamic simulations is expensive, we use an estimator of the final multiplicity based on the initial energy density $e(x,y,\eta_s)$ at mid-spacetime-rapidity~\cite{Giacalone:2019ldn}
\begin{equation}
\label{eq:estimator}
\frac{\dd N_\mathrm{ch}}{\dd\eta} = C\alpha\!\int\!\big[\tau e(x,y,\eta_s\!=\!0)\big]_0^{2/3}\,\dd x\,\dd y,
\end{equation}
with $\alpha = \frac{4}{3}(N_\mathrm{ch}/S) C_\infty^{3/4} (4\pi\eta/s)^{1/3} ( \pi^2\nu_\mathrm{eff}/30)^{1/3}$ the proportionality factor advocated in Ref.~\cite{Giacalone:2019ldn},\footnote{We use $S/N_\mathrm{ch} = 7.5$~\cite{Hanus:2019fnc}, $C_\infty = 0.87$~\cite{Giacalone:2019ldn}, and $\nu_\mathrm{eff} = 40$.} while $C$ is a constant of order 1.
Setting first $K_g=1$ in \McDipper\ and comparing the estimate from Eq.~\eqref{eq:estimator} and the outcome of the full \MUSIC+\iSS\ evolution for a small number of events, we determine the value $C=1.13$ to ``calibrate'' the estimator formula. 
Subsequently, a larger set of events is generated while varying $K_g$, until the multiplicity obtained from Eq.~\eqref{eq:estimator} matches the experimental data over the 0--60\% centrality range. 
The optimal value for this setup is found to be $K_g = 3.066$.\footnote{In Ref.~\cite{Garcia-Montero:2025bpn}, the value $K_g = 2.71$ is found to reproduce the multiplicity in central Pb+Pb collisions at $\sqrt{\snn} = 2.76$ TeV for the same \McDipper\ setup (but with a different subsequent evolution).}
The agreement between the multiplicities computed in our approach and preliminary ALICE data is shown in Fig.~\ref{fig:calibration}.

\begin{figure}
	\includegraphics[width=0.95\linewidth]{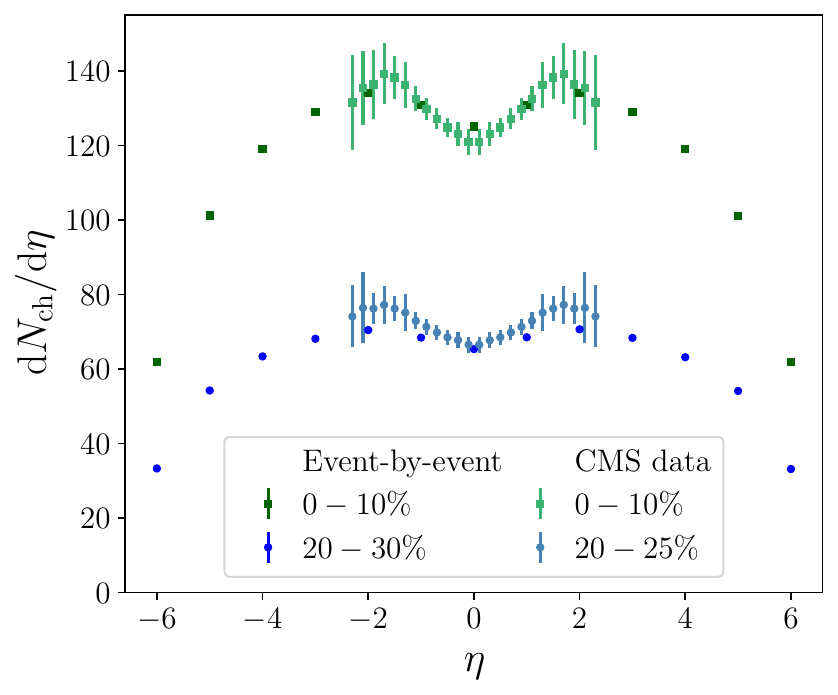}\vspace{-3mm}
	\caption{Pseudorapidity dependence of the charged-hadron multiplicity in O+O collisions at $\sqrt{\snn} = 5.36$ TeV from our model compared with CMS data (with systematic error bars only)~\cite{CMS:2026sai}.}
	\label{fig:Nch_vs_eta}
\end{figure}

\begin{figure*}
	\includegraphics[width=\linewidth]{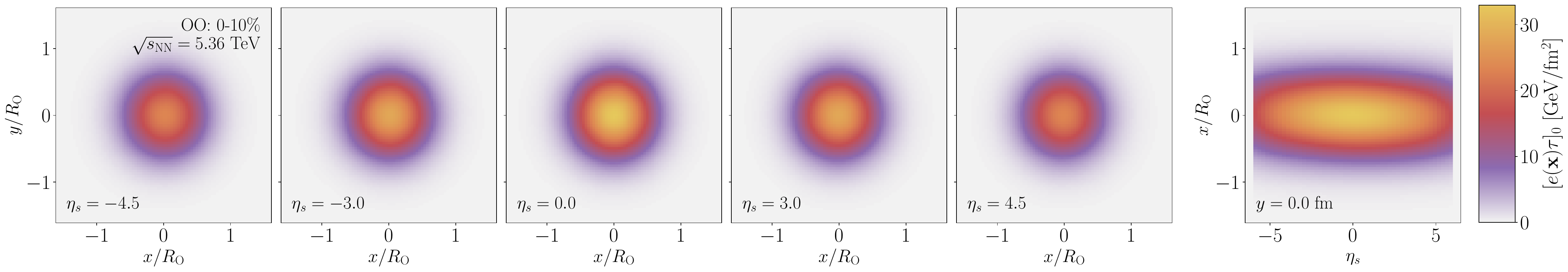}
	\includegraphics[width=\linewidth]{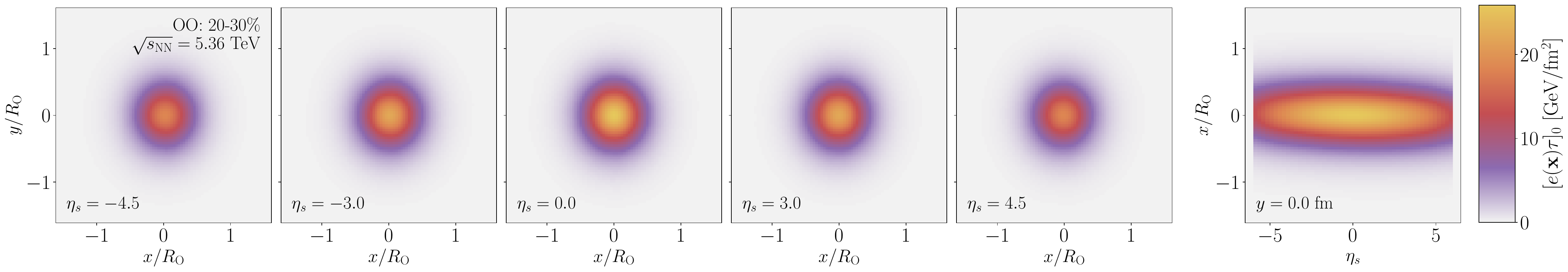}\vspace{-2mm}
	\caption{Energy profile for the average initial state in O+O collisions at $\sqrt{\snn} = 5.36$ TeV for 0--10\% (top) and 20--30\% (bottom) centrality. 
	For each centrality, the first five panels show transverse energy profiles in different spacetime rapidity windows, while the rightmost panel displays the energy in the $(x,\eta_s)$-plane for $y=0$ fm. 
	The $x$- and $y$-axes are in units of the radius $R_{\rm O} = 2.608$~fm used in the three-parameter-Fermi density distribution for ${}^{16}$O~\cite{Loizides:2014vua}.}
\label{fig:AvgState}
\end{figure*}

In the following, we shall only use the 0--10\% and the 20--30\% centrality classes. 
At lower multiplicities, the meaningfulness of applying hydrodynamics may become disputable~\cite{Constantin:2025ova,Ito:2026hec}.
In Fig.~\ref{fig:Nch_vs_eta}, we compare the prediction of our \McDipper+\MUSIC+\iSS\ model for the pseudorapidity dependence of charged multiplicity in those two classes to recent CMS data~\cite{CMS:2026sai}, where we average the results (and error bars) from their 0--5\% and 5--10\% classes to produce the 0--10\% class. 
Note that statistical error bars in the model, estimated from a jackknife, are too small to be seen.


\section{Results}
\label{sec:results}

In this section, we first discuss the properties of the average initial state generated in O+O collisions in our model (Sec.~\ref{sec:AvgState}). 
We then recall briefly the principle of the decomposition of initial-state fluctuations in uncorrelated modes and apply it to oxygen collisions in Sec.~\ref{sec:Modes}. 
Finally, we focus in Sec.~\ref{sec:v2e2} on elliptic flow: 
We show that the typical behavior of fluctuations can be captured by a few modes, but that a larger number is needed for a more complicated observable like the longitudinal decorrelation of eccentricity, which crucially depends on fluctuation modes that have a small relative contribution to the typical event.

\subsection{Average initial state}
\label{sec:AvgState}

From a large number $\Nev$ of random initial-state profiles $\{\Phi^{(i)}\}$ selected according to a given criterion, here the centrality class, one can define the average initial state $\bar{\Psi}$ as the arithmetic mean of the profiles:
\begin{equation}
\label{eq:av-state}
\bar{\Psi}(x, y, \eta_s) \equiv \frac{1}{N_\mathrm{ev}}\!\sum_{i=1}^{\Nev} \Phi^{(i)\!}(x, y, \eta_s) 
\equiv \langle \Phi(x, y, \eta_s) \rangle,
\end{equation}
where throughout this paper angular brackets denote an average over events.
Letting $\bar{\Psi}$ evolve with a given setup, one produces a fictitious event, whose final-state properties can be computed. 
In collisions of large nuclei, like Pb+Pb or Au+Au, the outcome of the evolution of $\bar{\Psi}$ is a good first-approximation proxy for the mean behavior observed in a sample of events for a number of global observables like multiplicity, mean transverse momentum, or elliptic and quadrangular flow~\cite{Borghini:2022iym, Krupczak:2025gwg}. 

We generated $\Nev = 2^{17}$ initial \McDipper\ profiles from O+O collisions in each centrality class of interest (0--10\% and 20--30\%), keeping the impact parameter direction along the $x$-axis.
The resulting average initial states are shown in Fig.~\ref{fig:AvgState}, in which we display the transverse dependence of the energy density in five different spacetime-rapidity windows, together with the energy profiles in the $(x,\eta_s)$-plane (at $y=0$) in the rightmost panels.

\begin{figure}[t]
	\includegraphics[width=0.9\linewidth]{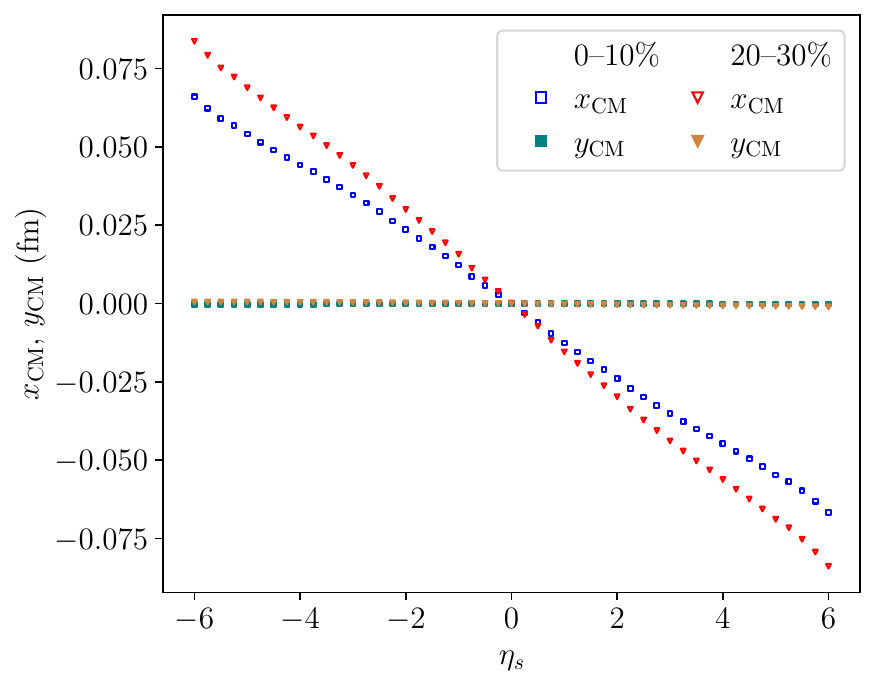}\vspace{-3mm}
	\caption{Spacetime rapidity dependence of the coordinates $(x_{\rm CM},y_{\rm CM})$ (in fm) of the center-of-mass of each $\eta_s$-slice of the average initial state $\bar{\Psi}$.}
	\label{fig:CM}
\end{figure}

As could be anticipated, these average initial states $\bar{\Psi}$ have smooth profiles. 
The highest energy-density value reached in the initial average state for the 0--10\% class is larger by about 27\% than in the 20--30\% class. 
Combined with the wider transversal extent, this leads to a larger transverse energy, which at the end of the evolution results in a factor of almost 2 in the charged hadron multiplicity at midrapidity: $\dd N_{\rm ch}/\dd\eta = 133$ vs.\ 69.

Since \McDipper\ is a 3D model, the energy density is not boost invariant, but its value at $x=y=0$ decreases by about half between midrapidity and $|\eta_s| = 5$. 
Although this is hardly visible in the rightmost panel of Fig.~\ref{fig:AvgState}, there is a small tilt of the system along $\eta_s$.
This can be quantified by looking at the position $(x_{\rm CM},y_{\rm CM})$ of the center-of-mass (weighted by the energy density) of each $\eta_s$-slice, as shown in Fig.~\ref{fig:CM}. 
As expected from the $y\to -y$ symmetry of the O+O collisions, no tilt is observed in the $y$ direction. 
In contrast, along the impact-parameter direction the $x$-coordinate of the center of mass exhibits a small rapidity-odd tilt. 
This deformation arises from the asymmetry between ``target'' and ``projectile'' away from $\eta_s=0$. 
At the high collision energy considered here the interaction between the nuclei as they fly through each other happens over a very short timescale, resulting in a relatively small tilt compared to lower-energy collisions, but it is still present.

Turning to the transverse profiles, a notable observation is that collisions of oxygen nuclei produce an average initial state which looks round-shaped at both centralities considered here. 
This is to be expected in ``central'' collisions, but not for events in the 20--30\% centrality class: for the latter, the average initial state in collisions of heavy nuclei is significantly more almond-shaped than here. 
Going beyond the visual impression, the asymmetry of the geometry can be quantified by the elliptic eccentricities $\varepsilon_{2,\rm c}(\eta_s)$ and $\varepsilon_{2,\rm s}(\eta_s)$ defined as the real and imaginary parts of
\begin{equation}
	\label{eq:eps2}
\varepsilon_{2,\rm c}(\eta_s) + \ii_{}\varepsilon_{2,\rm s}(\eta_s) \equiv -\frac{\displaystyle\int\!r^2\ee^{2\ii\theta}e(r,\theta,\eta_s)\, r\,\dd r\,\dd\theta}{\displaystyle\int\!r^{2\,}e(r,\theta,\eta_s)\, r\,\dd r\,\dd\theta},
\end{equation}
where $(r,\theta)$ are centered polar coordinates in the spacetime rapidity slice at $\eta_s$, using the center-of-mass of that slice~\cite{Shen:2020jwv}, and with the $x$-axis as origin of polar angles. 
The rapidity dependence of these eccentricities is displayed in Fig.~\ref{fig:e2_AvgState} and readily described: 
\begin{figure}[t]
	\includegraphics[width=0.9\linewidth]{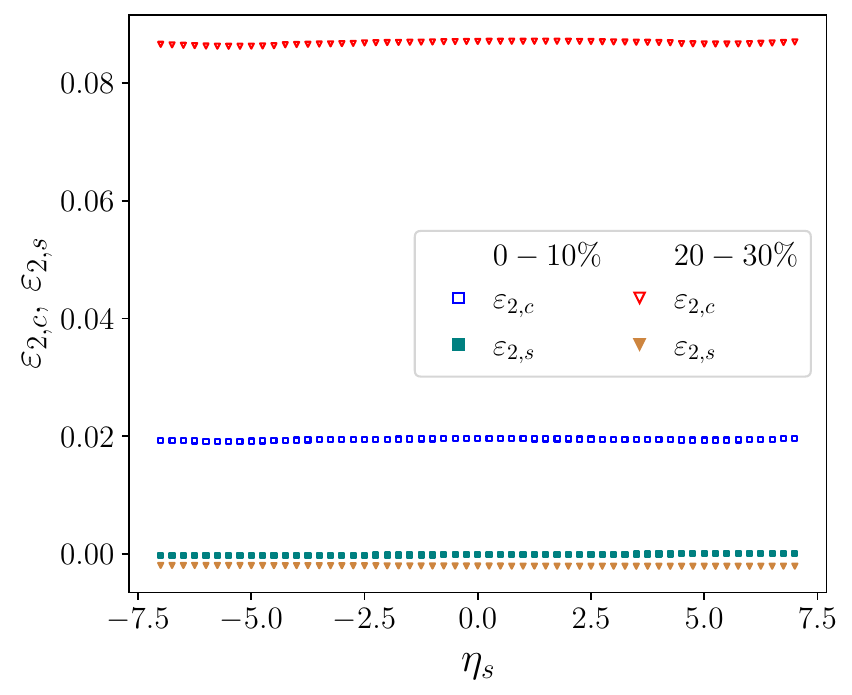}\vspace{-3mm}
	\caption{Spacetime rapidity dependence of the cosine and sine components of the elliptic eccentricity, Eq.~\eqref{eq:eps2}, for the average initial state $\bar{\Psi}$ in the 0--10\% and 20--30\% centrality classes.}
	\label{fig:e2_AvgState}
\end{figure}
At both centralities, the sine components $\varepsilon_{2,\rm s}$ are roughly zero across the rapidity range, again consistent with the $y\to -y$ symmetry of the colliding system. 
In turn, the cosine components $\varepsilon_{2,\rm c}$ are non-zero, but also almost constant over the $\eta_s$-range: $\varepsilon_{2,\rm c}\simeq 0.0195$ for the initial average state of the 0--10\% class, and $\varepsilon_{2,\rm c}\simeq 0.087$ in the 20--30\% class. 
For comparison, in Pb+Pb collisions at $\sqrt{\snn} = 5.02$~TeV (and using a Monte Carlo Glauber initial-state model) we found~\cite{Krupczak:2025gwg}  $\varepsilon_{2,\rm c}\simeq 0.017$ for 0--10\% central collisions, similar to the value in O+O collisions, but $\varepsilon_{2,\rm c}\simeq 0.32$ in the 30--40\% class: 
interpolating yields $\varepsilon_{2,\rm c}\gtrsim 0.2$ in 20--30\% Pb+Pb collisions, significantly larger than in O+O. 

The smallness of the ellipticity of the average initial state in O+O collisions is also manifest when one compares the values given above with those found in approaches that compute some ``average ellipticity'' of collisions in a centrality class or at a given impact-parameter value~\cite{Lim:2018huo,Sievert:2019zjr,Rybczynski:2019adt,Prasad:2024ahm,MenonKavumpadikkalRadhakrishnan:2025apq,Loizides:2025ule}.
In these studies, varying assumptions are used for the distribution of nucleons inside ${}^{16}$O nuclei or for the energy/entropy deposition scheme, which clearly affect the initial-state geometry. 
In turn, the elliptic shape of that geometry is characterized by various quantities, as e.g.\ the mean participant eccentricity or estimates based on two- or four-point cumulants of the energy (or entropy) distribution. 
Yet the generic finding is that the values of the event-averaged $\varepsilon_2$ are smallest in most central collisions (or at zero impact parameter), where they lie between 0.2 and 0.35. 
That is an order of magnitude larger than what we find for $\varepsilon_2\equiv(\varepsilon_{2,\rm c}^2+\varepsilon_{2,\rm s}^2)^{1/2}$ of the average initial state $\bar{\Psi}$ in central collisions. 
Similarly, the value $\varepsilon_2(\bar{\Psi})\simeq 0.087$ for the 20--30\% centrality range is a factor 4 to 5 smaller than values obtained from averaging over events. 
This does not signal that our model can already be discarded: as we shall see in Sec.~\ref{sec:v2e2}, we do find event-averaged values that are significantly larger than $\varepsilon_2(\bar{\Psi})$. 
What it means is that the value of $\varepsilon_2$ of the average initial state $\bar{\Psi}$ is not a good estimate of the mean value obtained by averaging the eccentricities of a set of events. 
In turn, it means that the average initial state alone is not sufficient to explain the elliptic flow, but that event-by-event fluctuations, which we discuss next, play a crucial role.

\subsection{Event-by-event fluctuations}
\label{sec:Modes}

Any real initial state $\Phi^{(i)}$ will differ from the average state $\bar{\Psi}$. 
\begin{figure}
	\includegraphics[width=\linewidth]{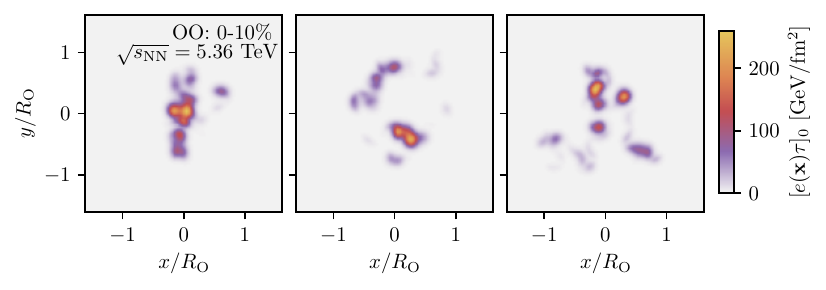}
	\includegraphics[width=\linewidth]{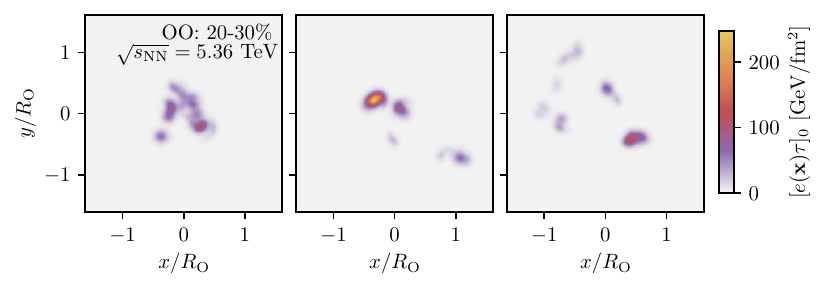}\vspace{-3mm}
	\caption{Transverse energy profile at $\eta_s=0$ of random O+O events at $\sqrt{\snn} = 5.36$ TeV belonging to the 0--10\% (top) and 20--30\% (bottom) centrality class. 
	Both axes are in units of $R_{\rm O} = 2.608$~fm.
	}
	\label{fig:random}
\end{figure}
As an example, we display in Fig.~\ref{fig:random} the transverse energy profiles at mid-spacetime-rapidity of three random O+O events for each centrality class, which are obviously quite different from the corresponding $\eta_s=0$ slices of $\bar{\Psi}$ in Fig.~\ref{fig:AvgState}. 
First, the energy density is not smoothly distributed, due to the small number of participant subnucleonic degrees of freedom in the colliding ${}^{16}$O nuclei. 
Second, the overall energy scale of Fig.~\ref{fig:random} differs by a factor of about 8 from that of Fig.~\ref{fig:AvgState}: 
Random O+O initial states may consist of a few localized high-energy spots, which altogether amount to a similar entropy as the colder, more extended profile of the average initial state, and thus yield --- within the dynamical model --- a similar final charged hadron multiplicity. 
Third, these random events can have a very asymmetric geometry, as e.g.\ the first initial state in the top row of Fig.~\ref{fig:random}, thus a ``central'' event, which clearly exhibits a highly ``elliptical'' shape.
That is, the definition of centrality via the charged multiplicity is not strongly correlated with geometric shape, as it is in collisions of heavy nuclei.
In fact, apart from the energy-density scale, there is no obvious difference between the shapes of the events in the top and bottom rows of Fig.~\ref{fig:random}.

From the previous remarks it is clear that the so-called ``event-by-event fluctuations'', namely the departure $\delta\Phi^{(i)}\equiv \Phi^{(i)}-\bar{\Psi}$ of each individual initial state from the average $\bar{\Psi}$ of the centrality class, will play a major role in geometry-related observables. 
In addition, these fluctuations are not small, as is the case in collisions of heavy nuclei, but can be large, as exemplified by the scale of the hot spots compared to that of the average $\bar{\Psi}$. 
That is, if one wishes to relate event-by-event fluctuations of final-state observables to corresponding fluctuations in the initial state, one cannot assume simple linear relations. 
A framework that allows to study systematically the response of observables to initial-state fluctuations at arbitrary order --- and at a moderate cost at least up to quadratic order --- is that introduced in Ref.~\cite{Borghini:2022iym}, which we now briefly present. 

From a set of initial state profiles $\{\Phi^{(i)}\}$, which hereafter will be 3D \McDipper\ energy-density distributions, one constructs an appropriate orthogonal basis of fluctuation modes $\{\Psi_l\}$, such that each individual initial state can be written as
\begin{equation}
\Phi^{(i)\!}(x, y, \eta_s) = \bar{\Psi}(x, y, \eta_s) + \sum_l c_l^{(i)} \Psi_l(x, y, \eta_s),
\label{eq:decomposition}
\end{equation}
where $\bar{\Psi}$ is the average initial state~\eqref{eq:av-state}, while the expansion coefficients $c_l^{(i)}$ satisfy two properties: 
Viewed as realization of random variables $\{c_l\}$, each of these is centered, $\langle c_l \rangle = 0$ for every $l$, and the covariance matrix of the variables is the identity matrix, $\langle c_l c_{l'} \rangle = \delta_{ll'}$. 
In particular, the variance of a given $c_l$ is equal to 1, so that the typical magnitude of any $c_l^{(i)}$ is also unity.

To ensure the property regarding the covariance, the fluctuation modes $\{\Psi_l\}$ are the eigenvectors of the autocorrelation matrix of the energy-density fluctuations
\begin{equation}
\label{eq:rho}
\rho(x, y, \eta_s; x', y', \eta'_s) \equiv \langle \delta\Phi(x,y,\eta_s)_{} \delta\Phi(x',y',\eta'_s)\rangle.
\end{equation}
The eigenvalues $\{\lambda_l\}$ of this matrix quantify the relative importance of each fluctuation mode, which is why we order the $\{\Psi_l\}$ by decreasing $\lambda_l$.
In fact, due the condition imposed on the unit variance of each $c_l$, the $\{\Psi_l\}$ cannot be normalized to unity. 
Instead, the eigenvalue $\lambda_l$ is precisely the squared norm $\norm{\Psi_l}^2$, defined as the integral of the square of $\Psi_l$ over the whole space. 

\begin{figure*}
    \includegraphics[width=0.9\linewidth]{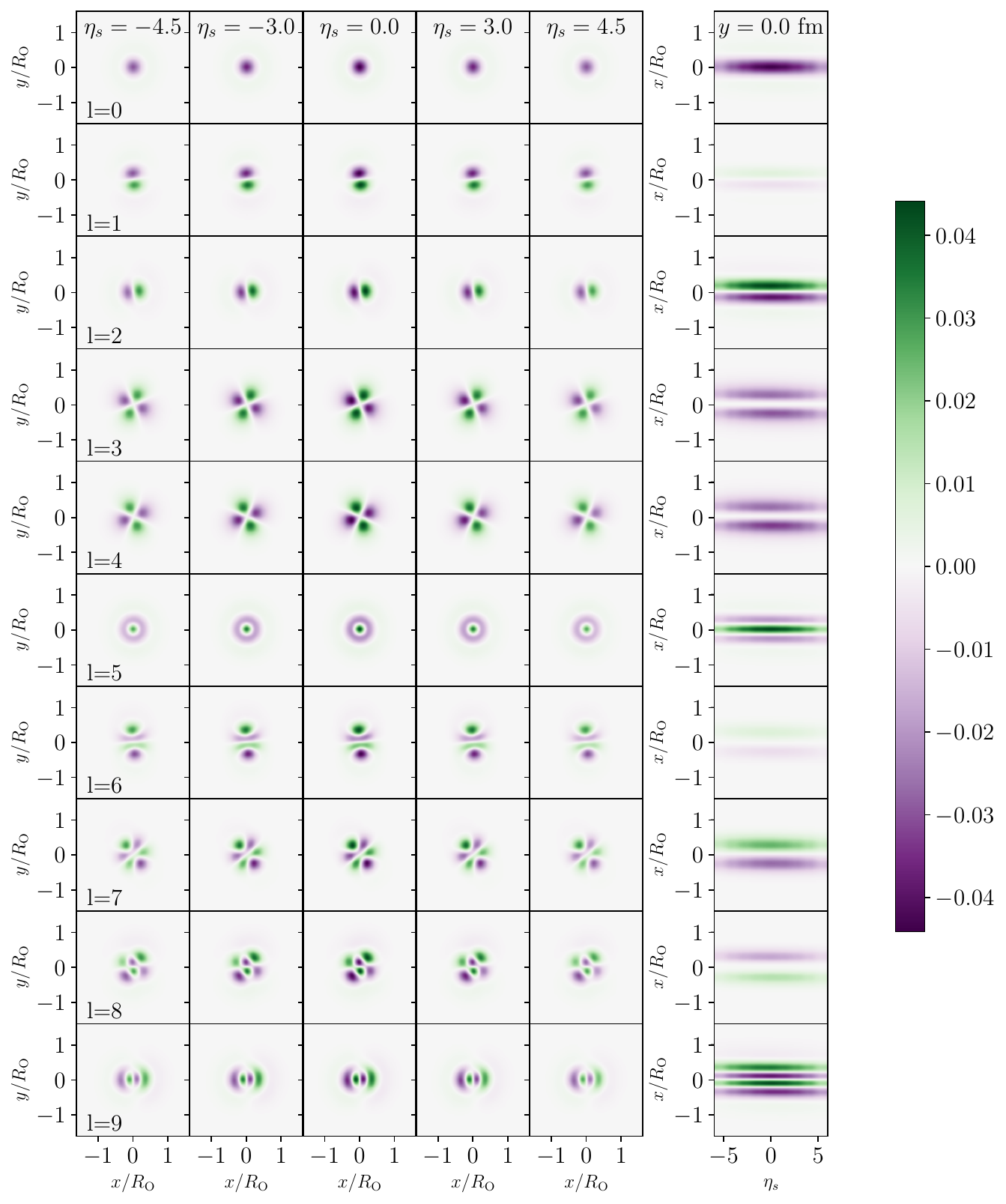}\vspace{-3mm}
    \caption{Normalized profiles of the first 10 fluctuation modes $\Psi_l$ for O+O collisions at 20--30\% centrality. 
	For each mode ($l=0$ to 9 from top to bottom), the first five panels show the transverse energy profiles in different $\eta_s$ windows, while the rightmost panel displays the profile in the $(x,\eta_s)$-plane for $y=0$~fm. 
    The $x$- and $y$-axes are in units of $R_{\rm O}=2.608$ fm.
    }
    \label{fig:10modes_20-30}
\end{figure*}

Previously, the decomposition in uncorrelated fluctuation modes has only been applied to two-dimensional initial-state profiles~\cite{Borghini:2022iym,Borghini:2024ekn,Krupczak:2025gwg}. 
Here we use it for the first time for 3D initial states.\footnote{Note that the autocorrelation matrix~\eqref{eq:rho} that determines the fluctuation modes is of size $(84\times 84\times 9)^2 = 63504\times 63504$, which is why we consider only 9 $\eta_s$-slices to compute the modes.}

\begin{figure*}
	\includegraphics[width=0.9\linewidth]{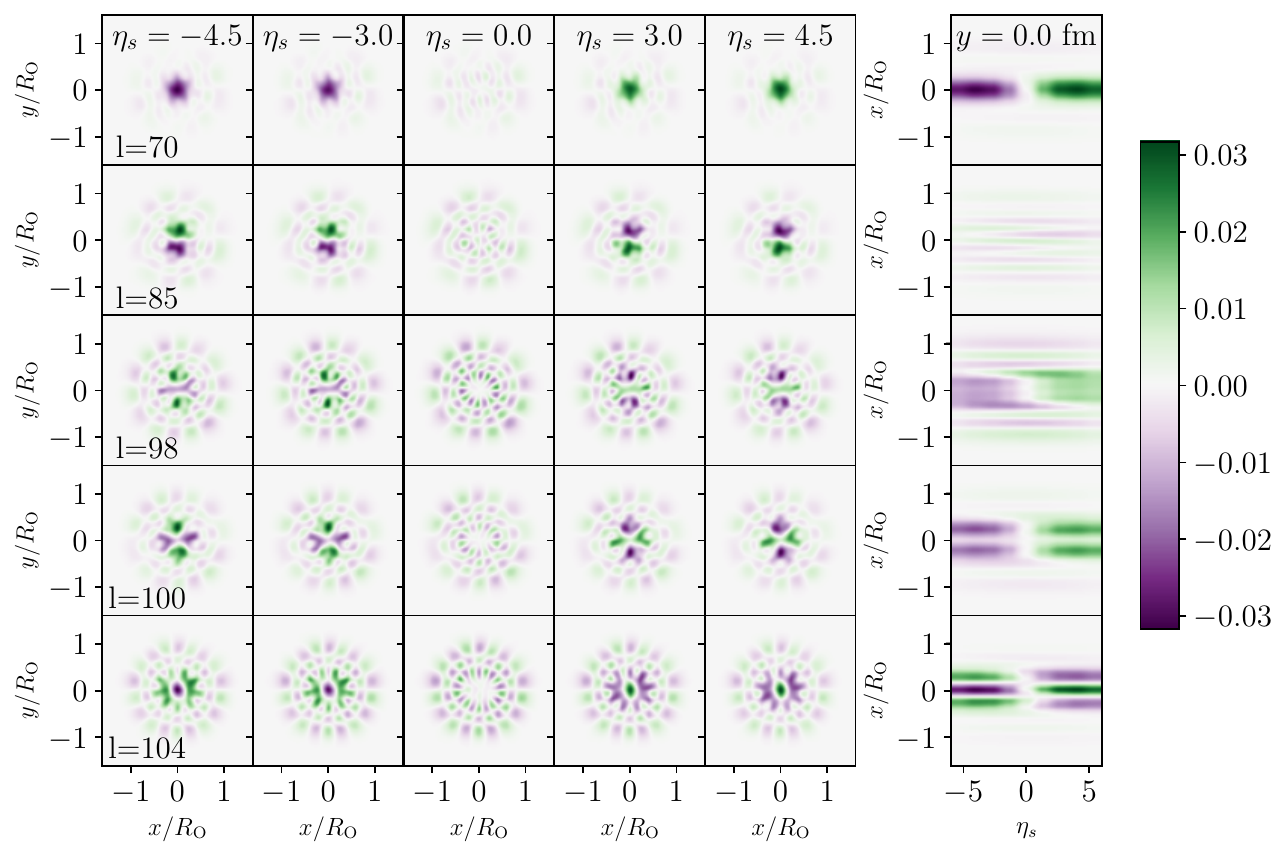}\vspace{-3mm}
	\caption{Same as Fig.~\ref{fig:10modes_20-30} for the first five fluctuation modes (from top to bottom: $l=70$, 85, 98, 100 and 104) with a significant spacetime-rapidity-odd component for O+O collisions in the 0--10\% centrality class.
	}
	\label{fig:odd_modes_0-10}
\end{figure*}

We start by displaying in Fig.~\ref{fig:10modes_20-30} the energy profiles of the first 10 modes, i.e.\ the ones with the largest associated $\lambda_l$, for non-central (20–30\%) collisions.\footnote{The profiles of the leading modes in central collisions are very similar and not shown.}
To allow comparison between the different modes, we do not display their absolute values, but the corresponding eigenvectors normalized to unity. 
For each mode, we show the same profiles as for the average initial state $\bar{\Psi}$ in Fig.~\ref{fig:AvgState}, namely the transverse energy in five different $\eta_s$ windows, and in the rightmost panel the longitudinal profile in the $(x,\eta_s)$-plane at $y=0$. 
Focusing first on the midrapidity slice $\eta_s=0$, one observes the same systematics as found for fluctuation modes of 2D initial states~\cite{Borghini:2022iym,Borghini:2024ekn,Krupczak:2025gwg}, namely modes with radial symmetry ($l=0$ and 5), with dipole structure (e.g.\ $l=1,2$, rotated by $90^{\rm o}$ with respect to each other), with quadrupole structure ($l=3,4$, rotated by $45^{\rm o}$), and more complex geometries. 
This can be characterized quantitatively in terms of eccentricities (and other observables), which we will do more systematically in a forthcoming paper. 
Here we just want to note that for these leading modes the structures present at midrapidity persist without significant change up to at least $\eta_s=\pm 4.5$, the only difference being that away from midrapidity less energy is deposited at a given transverse position $\bm{x}$ than at the same $\bm{x}$ at $\eta_s=0$. 
That is, the transverse geometry of these modes is approximately boost invariant, even though the energy content decreases when going to forward / backward regions. 
This relatively simple longitudinal behavior reflects the implementation of fluctuations in \McDipper, in which energy is deposited over regions with a small transverse size, but extend across a few units in spacetime rapidity (at fixed $\bm{x}$), as illustrated in Fig.~15 of Ref.~\cite{Garcia-Montero:2025bpn}.

Going to higher fluctuation modes, one finds a departure from this even dependence on $\eta_s$. 
Indeed, there appear modes that are odd in rapidity, or at least, with a significant $\eta_s$-odd component. 
This is illustrated in Fig.~\ref{fig:odd_modes_0-10}, which shows the five first (normalized) modes with a marked $\eta_s$-odd behavior for initial states in the 0--10\% centrality class. 
Thus, mode $l=70$ represents a fluctuation with a single roundish ``hot spot'' at the center of the transverse plane, but with a change of sign between positive and negative $\eta_s$ --- and accordingly almost no energy content at midrapidity. 
The next four modes are ``contaminated'' by a complicated even component, but with clearly identifiable $\eta_s$-odd structures. 
Thus, mode $l=85$ exhibits a dipole along the $y$-direction, which is rotated by 180$^{\rm o}$ between negative and positive $\eta_s$. 
In turn mode $l=100$ shows a quadrupole, which changes sign between forward and backward rapidity. 
Mode $l=104$ (bottom row of Fig.~\ref{fig:odd_modes_0-10}) is the superposition of an even high-order multipole --- which would yield a sizable $\varepsilon_9$ --- and a simple rapidity-odd pattern in the center, which dominates the profile in the $(x,\eta_s)$-plane. 

\begin{figure*}
	\includegraphics[width=0.9\linewidth]{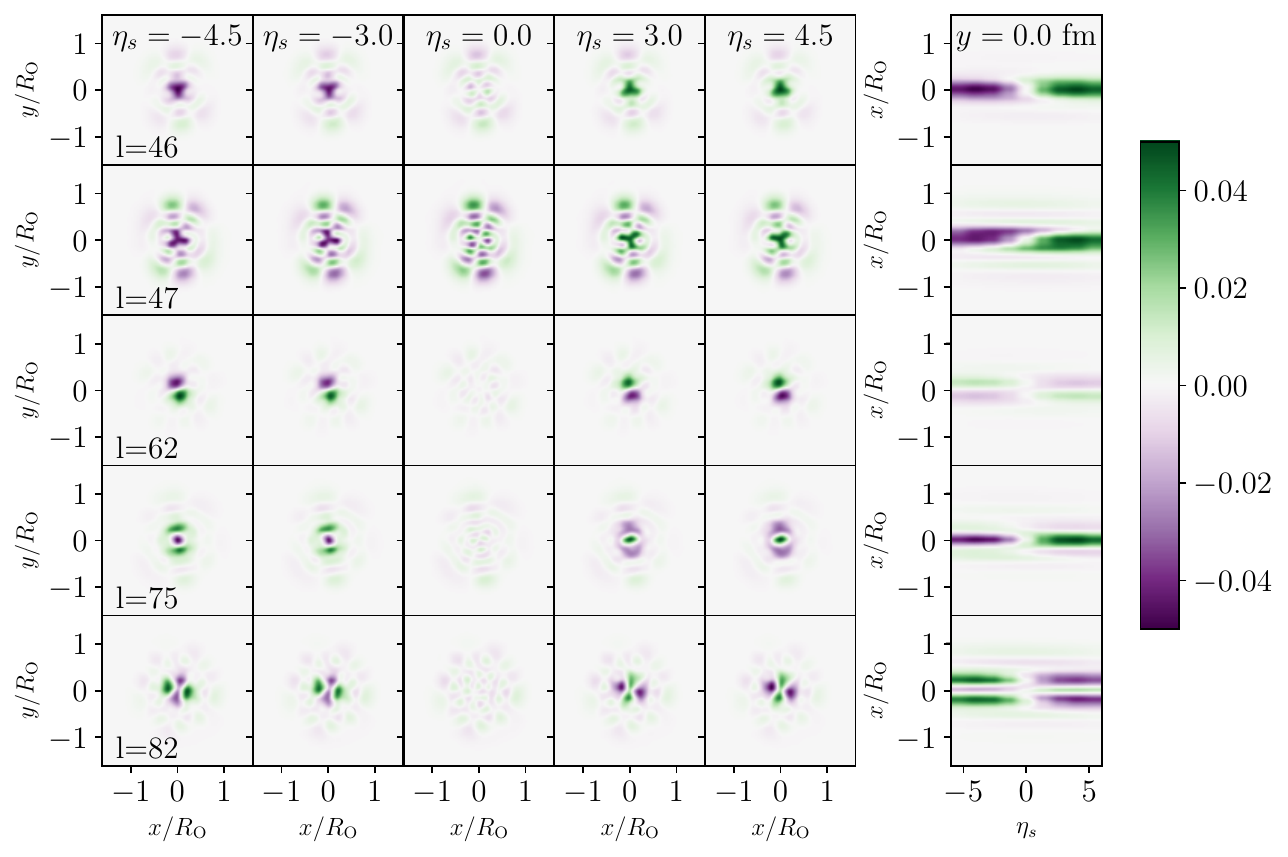}\vspace{-3mm}
	\caption{Same as Fig.~\ref{fig:odd_modes_0-10} for the first five fluctuation modes (from top to bottom: $l=46$, $47$, $62$, $75$ and $82$) with a significant spacetime-rapidity-odd component for O+O collisions in the 20--30\% centrality class.
	}
	\label{fig:odd_modes_20-30}
\end{figure*}

The first five modes with a large $\eta_s$-odd component in 20--30\%-central collisions (Fig.~\ref{fig:odd_modes_20-30}) have similar geometries, which we do not describe in detail. 
There is however a point worth mentioning, namely that these first (partly) odd modes are now those with $l=46$, 47, 62, 75 and 82: 
This is significantly earlier than in central events, and suggests that the fluctuation modes with an odd component are more important in the 20--30\% class. 
Since such modes play a crucial role in generating the so-called decorrelation of observables across different longitudinal slices, as discussed for elliptic flow in the next section, this is a hint that longitudinal decorrelation may be larger in non-central events.

To quantify the relative importance of the fluctuation modes, one can define weights for each mode and for the average state~\cite{Borghini:2022iym, Borghini:2024ekn}:%
\begin{equation}
    w_l \equiv \frac{\norm{\Psi_l}}{\sum_l \norm{\Psi_l} + \norm{\bar{\Psi}}}
    \quad\mathrm{and}\quad
    \bar{w} \equiv \frac{\norm{\bar{\Psi}}}{\sum_l \norm{\Psi_l} + \norm{\bar{\Psi}}},
\label{eq:weights}
\end{equation}
where we recall that the norm of a mode is given by the associated eigenvalue of the matrix~\eqref{eq:rho}: $\norm{\Psi_l}=\sqrt{\lambda_l}$. 
\begin{figure}[b]
	\includegraphics[width=0.9\linewidth]{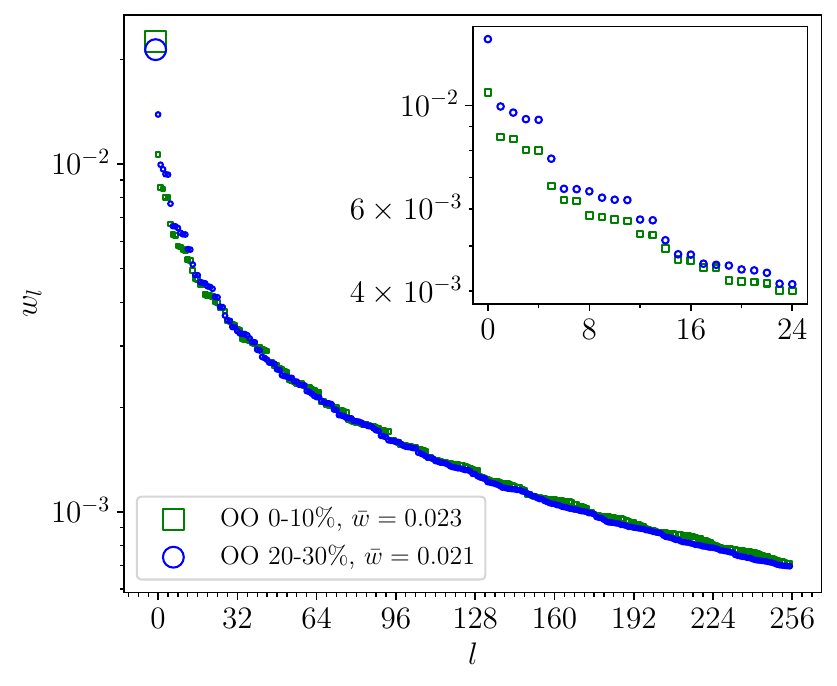}\vspace{-3mm}
	\caption{Relative weights~\eqref{eq:weights} of the average initial state (larger symbols at $l=-1$) and the first 256 fluctuation modes $\{\Psi_l\}$ for both centrality classes.}
	\label{fig:weights}
\end{figure}
These relative weights, computed using the largest $10^4$ eigenvalues in the sum in the denominator, are shown in Fig.~\ref{fig:weights} for the first 256 modes of both centrality classes, together with the weight $\bar{w}$ of the respective average states represented by the larger symbols at $l = -1$. 
Notably, the relative weight of the average state is only around 2\% at both centralities, and only larger than the weight $w_0$ of the leading mode $\Psi_0$ by about a factor 2 or even less. 
That is significantly smaller than in Pb+Pb events, in which $\bar{w}$ is about 13\% and a factor 4 or more larger than $w_0$~\cite{Krupczak:2025gwg}.
This illustrates quantitatively the idea that fluctuations are relatively more important in a smaller system like O+O than in Pb+Pb collisions. 

A second marked difference between Fig.~\ref{fig:weights} and the behavior in Pb+Pb collisions regards the comparison between the two centralities. 
In Pb+Pb initial states~\cite{Krupczak:2025gwg}, the spectrum of weights $\{w_l\}$ is flatter in central than in non-central collisions. 
In contrast, here both spectra decrease with $l$ at the same rate. 
There are small differences, in particular more (quasi)degenerate weights in central than in 20--30\% collisions, especially visible in the inset focusing on the first 25 modes. 
But the overall trend is that $w_l$ is quite similar at both centralities. 
Coming back to the discussion of Figs.~\ref{fig:odd_modes_0-10} and \ref{fig:odd_modes_20-30}, this means that $\eta_s$-odd modes have larger relative weights in 20--30\% collisions than in central ones, and thus they should have a stronger impact on observables. 

\subsection{Capturing fluctuations with a small set of modes}
\label{sec:v2e2}

From the preceding subsections it is clear that event-by-event fluctuations play an important role in O+O collisions, more so than in Pb+Pb, and are a necessary ingredient besides the average initial state for the description of observables. 
We now address the next question: how many modes are necessary to describe an observable?
Here we focus on elliptic flow $v_2$, and on the corresponding initial-state eccentricity $\varepsilon_2$, only. 

On the one hand, we generated a random sample of 5000 \McDipper\ initial states for each centrality, which constitute the basis for our ``event-by-event data'' in each class. 
From these, we determined the initial eccentricities $\varepsilon_{2,\rm c}(\eta_s)$ and $\varepsilon_{2,\rm s}(\eta_s)$ in each spacetime-rapidity bin. 
After the full hydrodynamic evolution using \MUSIC\ and particlization (including decays)  with \iSS, we computed for each event the (pseudorapidity-dependent) elliptic flow coefficients $v_{2,\rm c}(\eta)$ and $v_{2,\rm s}(\eta)$. 
The latter are respectively the real and imaginary part of the average value of $\ee^{2\ii_{}\varphi_{\bm{p}}}$, with $\varphi_{\bm{p}}$ the azimuth of the transverse momentum with respect to the $x$-axis. 

On the other hand, we generated sets of ``linearized observables'', following the recipe that we now detail. 
The starting point is that within the mode-by-mode-decomposition framework, an appropriate observable $O_\alpha$ can be computed in terms of response coefficients, which quantify how much a given fluctuation mode affects the observable~\cite{Borghini:2022iym}. 
At linear order, one can thus write
 \begin{equation}
    O_\alpha\bigg(\bar{\Psi} + \sum_l c_l\Psi_l\bigg) \simeq \bar{O}_\alpha + \sum_l c_l L_{\alpha,l}\,,
\label{eq:Obs_linear}
\end{equation}
where $\bar{O}_\alpha \equiv O_\alpha(\bar{\Psi})$ while the $\{L_{\alpha,l}\}$ are linear response coefficients. 
In practice, a given $L_{\alpha,l}$ --- or more precisely, in the case of a rapidity-dependent observable, a set of values of $L_{\alpha,l}$ for different $\eta_s$ or $\eta$ bins --- is computed by looking at artificial events with an initial state consisting of $\bar{\Psi}$ and a small component along the single mode $\Psi_l$, which yields at once the response for initial state quantities. 
Letting the artificial initial states evolve dynamically, one also obtains the response coefficients for final state observables. 

Thanks to the absence of correlation between modes, $\langle c_l c_{l'}\rangle = \delta_{ll'}$, the covariance of the fluctuations of two observables (or the variance of a single observable) is to a good approximation controlled by the linear response coefficients: 
$\langle (O_\alpha - \langle O_\alpha\rangle)(O_\beta - \langle O_\beta\rangle) \rangle \simeq \sum_l L_{\alpha,l}L_{\beta,l}$, where extra terms only appear when pushing expansion~\eqref{eq:Obs_linear} to third order in the $\{c_l\}$. 
Observables that are approximately linear in the sense of Eq.~\eqref{eq:Obs_linear} are initial eccentricities like $\varepsilon_{2,\rm c}$ or $\varepsilon_{2,\rm s}$, Eq.~\eqref{eq:eps2}.\footnote{In contrast, $\varepsilon_2\equiv\sqrt{\varepsilon_{2,\rm c}^2+\varepsilon_{2,\rm s}^2}$ is very nonlinear} 
If these are small, then $v_{2,\rm c}$ and $v_{2,\rm s}$ are almost proportional to $\varepsilon_{2,\rm c}$ and $\varepsilon_{2,\rm s}$, respectively, i.e.\ the flow coefficients are approximately linear observables in the sense of Eq.~\eqref{eq:Obs_linear}.
Yet one can anticipate that this holds to lesser extent than for the eccentricities, due to the nonlinearities introduced by the hydrodynamic evolution. 

Within the approximation~\eqref{eq:Obs_linear}, we created ``linearized observables'' as follows. 
We consider simplified initial states of the form~\eqref{eq:decomposition} including only a small number $N_{\rm modes}$ of fluctuation modes, namely those with the largest linear-response coefficients for $\varepsilon_{2,\rm c}$ or $\varepsilon_{2,\rm s}$ (up to some $l_{\max}$). 
For each mock event, $N_{\rm modes}$ coefficients $c_l$ are drawn randomly from a centered Gaussian distribution with unit variance\footnote{We checked that, for the modes under consideration, this is a good approximation to the distribution of $c_l$ which can be reconstructed by expanding the 5000 initial profiles of our event-by-event data according to Eq.~\eqref{eq:decomposition}. This is not true for all modes, in particular for the leading one $l=0$, which is significantly non-Gaussian and will be discussed in a future work.} and associated to the modes used in the construction. 
Eventually, we use the linear approximation~\eqref{eq:Obs_linear} to assign observables ($\varepsilon_{2,\rm c}$, $\varepsilon_{2,\rm s}$, $v_{2,\rm c}$ and $v_{2,\rm s}$) to the fake event. 
This in particular means that the elliptic flow coefficients are estimated without evolving dynamically the simplified initial states (which are actually never generated). 
Another consequence of using the linear approximation~\eqref{eq:Obs_linear} is that the absolute value of $\varepsilon_{2,\rm c}$ or $\varepsilon_{2,\rm s}$ can become larger than 1, which is non-physical. 
The actual eccentricities of the initial states are of course smaller than 1 in absolute value, thanks to nonlinear terms that go beyond Eq.~\eqref{eq:Obs_linear}, but are not taken into account in our linearized observables.
In total, we generated $5\times 10^4$ sets of linear $\varepsilon_{2,\rm c}(\eta_s)$, $\varepsilon_{2,\rm s}(\eta_s)$, $v_{2,\rm c}(\eta)$, $v_{2,\rm s}(\eta)$.
The corresponding results are labeled ``linear model'' in the figures.

For this set of linearized observables constructed from Gaussian-distributed fluctuations, one can actually write down the joint probability distribution, which is itself Gaussian~\cite{Borghini:2022iym}. 
For a set of $d$ observables, one has
\begin{equation}
    p(\{O_{\alpha_k}\}) = \frac{\exp\!\big[\!-\!\frac{1}{2} (O_{\alpha_i}\! - \bar{O}_{\alpha_i}) (\Sigma^{-1})_{ij} (O_{\alpha_j}\! - \bar{O}_{\alpha_j}) \big]}{(2\pi)^{d/2} \sqrt{\det \Sigma}},
\label{eq:joint_probability}
\end{equation}
where $\Sigma^{-1}$ is the inverse of the covariance matrix
\begin{equation}
    (\Sigma)_{ij} = \sum_l L_{\alpha_i,l} L_{\alpha_j,l}.
\end{equation}
For the sake of consistency, the sum runs over the same modes as those taken into account when constructing the linearized observables. 
The observables included in the joint distribution are the sine and cosine components of the initial eccentricities at the same spacetime rapidity and of elliptic flow coefficients at a given pseudorapidity, i.e.\ we consider $p(\varepsilon_{2,\rm c}(\eta_s), \varepsilon_{2,\rm s}(\eta_s), v_{2,\rm c}(\eta), v_{2,\rm s}(\eta))$. 
In this work, we restricted ourselves to the case $\eta = \eta_s$.
Results from this approach will be referred to as ``analytical G.f.a." (for Gaussian fluctuation approximation).

\begin{figure*}
	\includegraphics[width=0.495\linewidth]{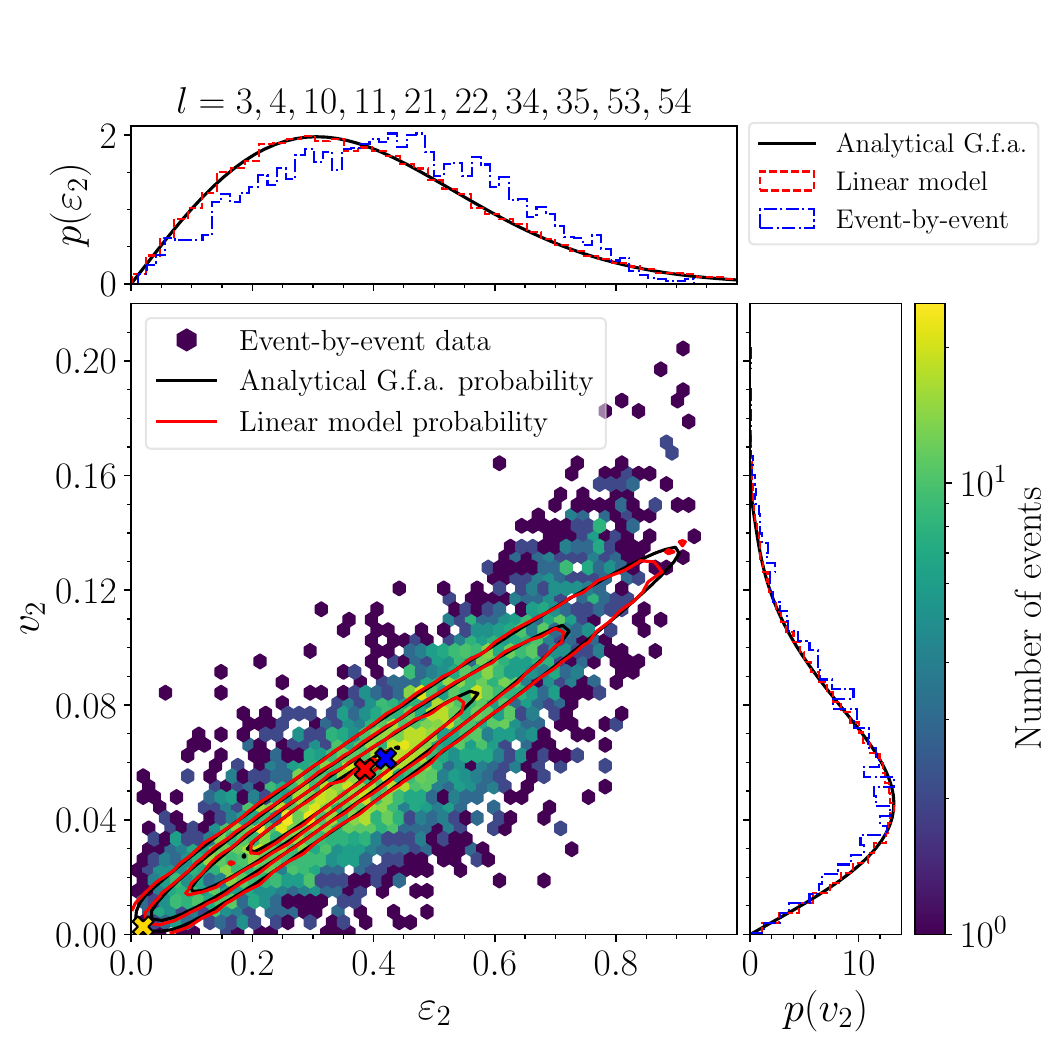}
	\includegraphics[width=0.495\linewidth]{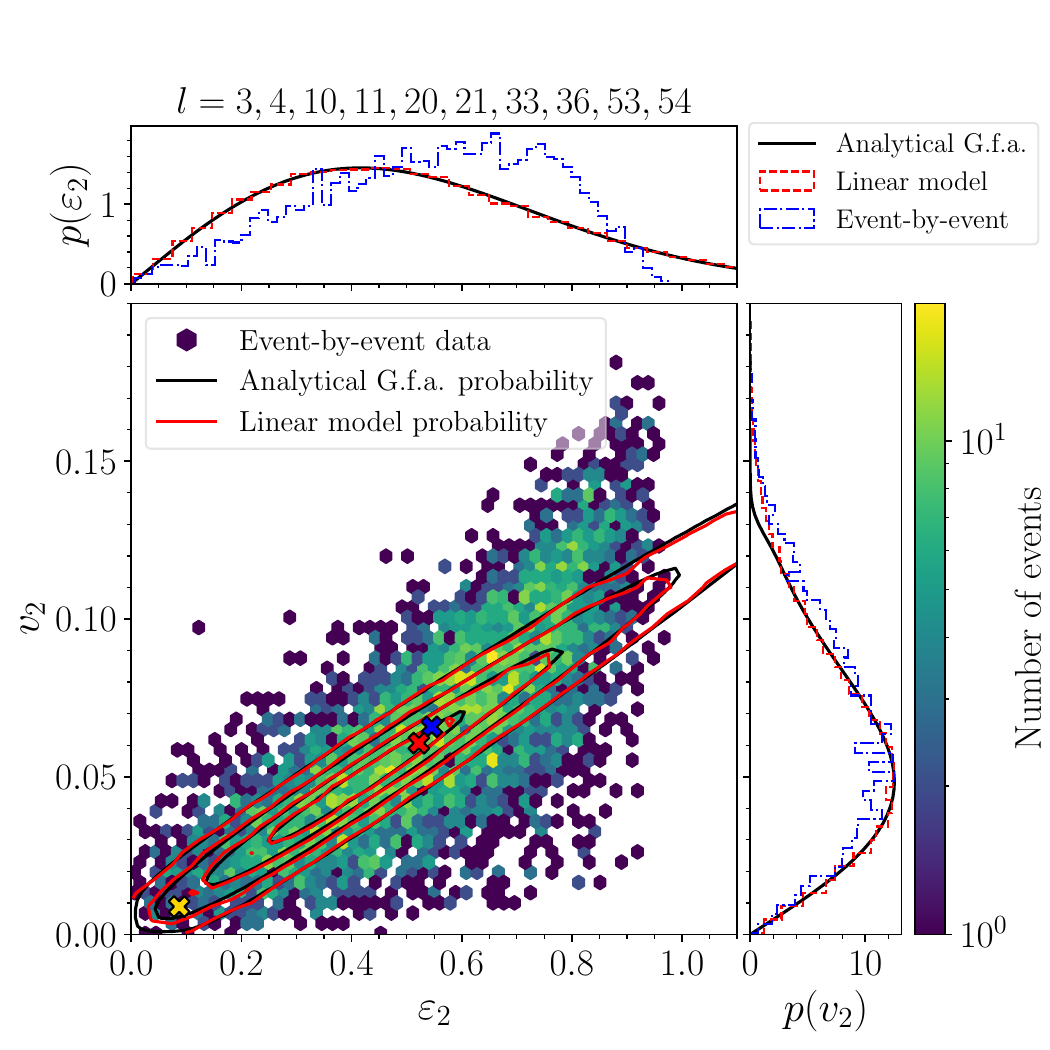}\vspace{-4mm}
	\caption{Joint probability distribution of $\varepsilon_2$ and $v_2$ at midrapidity in O+O collisions at 0--10\% (left) and 20--30\% (right) centrality. The smaller panels above and on the right show the marginal distributions $p(\varepsilon_2)$ and $p(v_2)$, respectively. 
		The hexagonal markers and the blue histograms represent event-by-event data from 5000 events, with the blue cross marking the average values of the observables. 
		The black lines correspond to the analytical probability from Eq.~\eqref{eq:joint_probability}, computed with 10 fluctuation modes (specified above each plot). The contour lines define regions that contain 20\%, 50\%, 80\%, and 95\% of the total probability. 
		The red curves and contours are obtained from $5\times 10^4$ sets of ``linearized observables'' $\varepsilon_{2,\rm c}$, $\varepsilon_{2,\rm s}$, $v_{2,\rm c}$, $v_{2,\rm s}$ created using Eq.~\eqref{eq:Obs_linear} with the same 10 modes and with Gaussian sampling of the coefficients $\{c_l\}$.
		The red cross marks the corresponding average value.
		The yellow cross is for the values of $\varepsilon_2$ and $v_2$ in the average state $\bar{\Psi}$.
		}
	\label{fig:v2_e2_10modes}
\end{figure*}

Figure~\ref{fig:v2_e2_10modes} represents as a heat map the joint probability $p(\varepsilon_2,v_2)$ of $\varepsilon_2$ and $v_2\equiv(v_{2,\rm c}^2+v_{2,\rm s}^2)^{1/2}$, both at midrapidity for our ``event-by-event data'' in both 0--10\% (left) and 20--30\% (right) classes.
From our larger sample of ``linear observables'' and from the Gaussian probability distribution~\eqref{eq:joint_probability}, both computed with 10 fluctuation modes only, we determined $p(\varepsilon_2,v_2)$ at midrapidity, for which respective contour lines are also shown in Fig.~\ref{fig:v2_e2_10modes}. 
The figure also displays the marginal distributions $p(\varepsilon_2)$ and $p(v_2)$ within the three approaches.\footnote{Similar plots at other rapidities give almost identical results, the only difference being a small change in the $v_2$-scale.} 

As expected, the ``linear model" and ``analytical G.f.a." results are in almost perfect agreement, since both rely on the same assumptions (validity of Eq.~\eqref{eq:Obs_linear} and Gaussian fluctuations) and are based on the same reduced set of modes.
In turn, they are in rather fair agreement with the event-by-event results for $p(\varepsilon_2,v_2)$, at least as long as $\varepsilon_2$ is smaller than about 0.6. 
At larger eccentricities, the values of $v_2$ at a given $\varepsilon_2$ are larger in the event-by-event data than in the calculations that ultimately rely on a linear relation between eccentricities and flow coefficients, which we shall further discuss hereafter. 
Coming back to $\varepsilon_2\lesssim 0.6$, we checked that by including more modes in the linear approaches, which raises the computational cost, one obtains an improved agreement with the ``real'' data: this mostly results in an broadening of the displayed contours, including rarer events into the description, but their slope is almost unchanged.

At large $\varepsilon_2$, the approaches based on linearly-responding observables depart from event-by-event data. 
This discrepancy is due to the ``banana'' shape of the latter, caused by the nonlinear dynamics relating the final state of the evolution to the initial condition. 
In particular, for elliptic flow one can write~\cite{Niemi:2015qia,Noronha-Hostler:2015dbi,Borghini:2018xum}
\begin{equation}
	v_2 \simeq \kappa_{2,2} \varepsilon_2 + \kappa_{2,222} \varepsilon_2^3, 
	\label{eq:kappa}
\end{equation} 
where the cubic term, which cannot be captured in a linear approximation, only becomes relevant for large $\varepsilon_2$. 
Here there are such large values of the eccentricity, almost up to 1, and even for central collisions --- remember the first event in the top row of Fig.~\ref{fig:random} ---, which leads to a sizable deviation from the linear behavior. 
At first sight, this contradicts the results of Ref.~\cite{Sievert:2019zjr}, in which the cubic term is found to be less important in O+O collisions than in larger systems. 
Yet one should note that between the present study and Ref.~\cite{Sievert:2019zjr} there are differences in the assumptions for the nuclear structure (NLEFT configurations vs.\ Woods-Saxon-distributed nucleons), for the initial-state model (smaller sub-nucleonic degrees of freedom within \McDipper\ vs.\ larger hot spots in $\mathrm{T}_{\mathrm{R}} \mathrm{ENTo}$), or even in the definitions of eccentricity used. 
All in all, in Ref.~\cite{Sievert:2019zjr} $\varepsilon_2\{2\}$ remains smaller than 0.5, in which regime we would also ignore the cubic response. 

Another issue at large $\varepsilon_2$ is that our linear model, in which eccentricity is not computed from real initial states but from the approximation~\eqref{eq:Obs_linear}, there do appear a number of cases in which $\varepsilon_2$ is larger than 1 in the 20--30\% centrality class, which of course never happens for real events. 
We could have taken this into account for our ``linear model'' data, by associating to each event its real  $\varepsilon_2$ computed from the initial state, instead of the value given by Eq.~\eqref{eq:Obs_linear}. 
This would indeed improve the agreement between the ``linear model'' and event-by-event results, but spoil the agreement with the analytical G.f.a.\ results.
A more consistent approach would be to include the quadratic order in the eccentricities, which is easy, and in the flow coefficients, which is computationally demanding, since an accurate determination of the quadratic-response coefficients requires high statistics to suppress statistical noise~\cite{Krupczak:2025gwg}. 
In parallel, the Gaussian-fluctuation approximation would need to be improved as well. 

The unrealistic $\varepsilon_2$-distribution and the absence of cubic response in the linear models strongly affect the marginal probability $p(\varepsilon_2)$, which is shifted to larger values, but with a quick cut when going towards 1, in event-by-event data. 
The discrepancy is less marked in $p(v_2)$, but there the linear models clearly miss the broader tail at large $v_2$ caused by the cubic response.

To be more quantitative, we present in Table~\ref{tab:e2_v2_avg} the sample mean of $\varepsilon_2$ and $v_2$ for the event-by-event data and for the values from the linear model. 
They are in reasonable agreement, differing by less than 10\%, at both centralities, at mid- and forward rapidity.
We checked that including more fluctuation modes in the linear model increases the agreement: 
including all 128 first modes, the event-by-event results are reproduced at the 1\% level.
Note that our $\langle\varepsilon_2\rangle$ values are in the same range than in previous theoretical studies~\cite{Lim:2018huo,Sievert:2019zjr,Rybczynski:2019adt,MenonKavumpadikkalRadhakrishnan:2025apq,Loizides:2025ule}, although we repeat that the models used differ. 
In turn, our values of $\langle v_2\rangle$ are comparable to experimental results for integrated elliptic flow~\cite{ALICE:2025luc,ATLAS:2025nnt,CMS:2025tga}, although care should again be observed, since the $v_2$-observables used differ, while centrality classes or the transverse-momentum ($p_{\rm T}$) ranges may not match --- our results are integrated over all possible $p_{\rm T}$.

We also give in Table~\ref{tab:e2_v2_avg} the values of $\varepsilon_2$ and $v_2$ computed for the average state. 
They deviate significantly, by about an order of magnitude, from the event-by-event averages. 
This contrasts with large systems, where the average state already provides a good approximation of the average over events~\cite{Krupczak:2025gwg}, due to the dominant geometric contribution.
That is, elliptic flow in O+O collisions is dominated by event-by-event fluctuations, as is triangular flow in larger (symmetric) systems. 
At the same time, the mean $v_2$ value can be accurately reproduced in our mode-by-mode approach by including only a limited number of fluctuation modes, which is a nice finding.

\begin{table}[t]
	\vspace{-2mm}
	\caption{\label{tab:e2_v2_avg}Average-event value $O(\bar{\Psi})$ and sample mean $\langle O\rangle$ of  $\varepsilon_2$ and $v_2$ for different centralities and longitudinal rapidities. 
	The sample mean is calculated from 5000 event-by-event simulations (E-by-E) or from the average of the $5\times 10^4$ values generated in the ``linear model'' using 10 modes.
	}
	\begin{ruledtabular}
		\begin{tabular}{cccccc}
			\multirow{2}{*}{centrality} & \multirow{2}{*}{$\eta_s$ or $\eta$} & \multirow{2}{*}{observable} & \multirow{2}{*}{$O(\bar{\Psi})$} & \multicolumn{2}{c}{sample mean} \\
			& & & & linear model\ & E-by-E \\
			\hline
			\multirow{2}{*}{0--10\%} & \multirow{2}{*}{$0$} 
			& $\varepsilon_2$ & 0.0196  & 0.3848 & 0.4199 \\
			&  & $ v_2$         & 0.0027  & 0.0574 & 0.0614 \\
			\hline
			\multirow{2}{*}{0--10\%} & \multirow{2}{*}{$4$} 
			& $\varepsilon_2$ & 0.0194 & 0.3768 & 0.4181 \\
			&  & $v_2$         & 0.0022 & 0.0472 & 0.0526 \\
			\hline
			\multirow{2}{*}{20--30\%} & \multirow{2}{*}{$0$} 
			& $\varepsilon_2$ & 0.0871 & 0.5200 & 0.5456 \\
			&  & $v_2$         & 0.0088 & 0.0605 & 0.0658 \\
			\hline
			\multirow{2}{*}{20--30\%} & \multirow{2}{*}{$4$} 
			& $\varepsilon_2$ & 0.0869 & 0.5142 & 0.5430 \\
			&  & $v_2$         & 0.0076 & 0.0500 & 0.0559 \\
		\end{tabular}
	\end{ruledtabular}
\end{table}

\begin{figure}
	\includegraphics[width=0.85\linewidth]{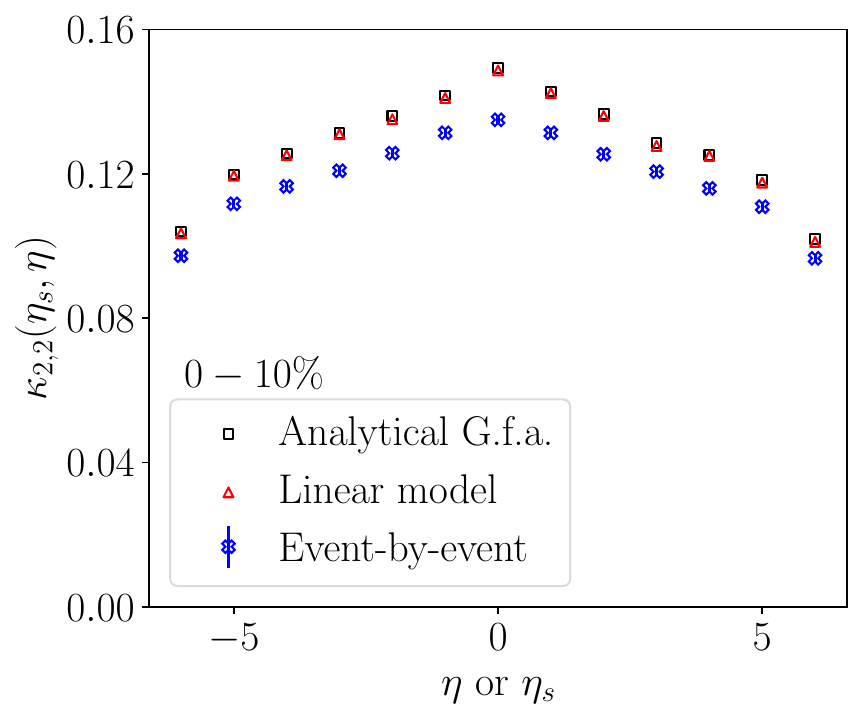}
	\includegraphics[width=0.85\linewidth]{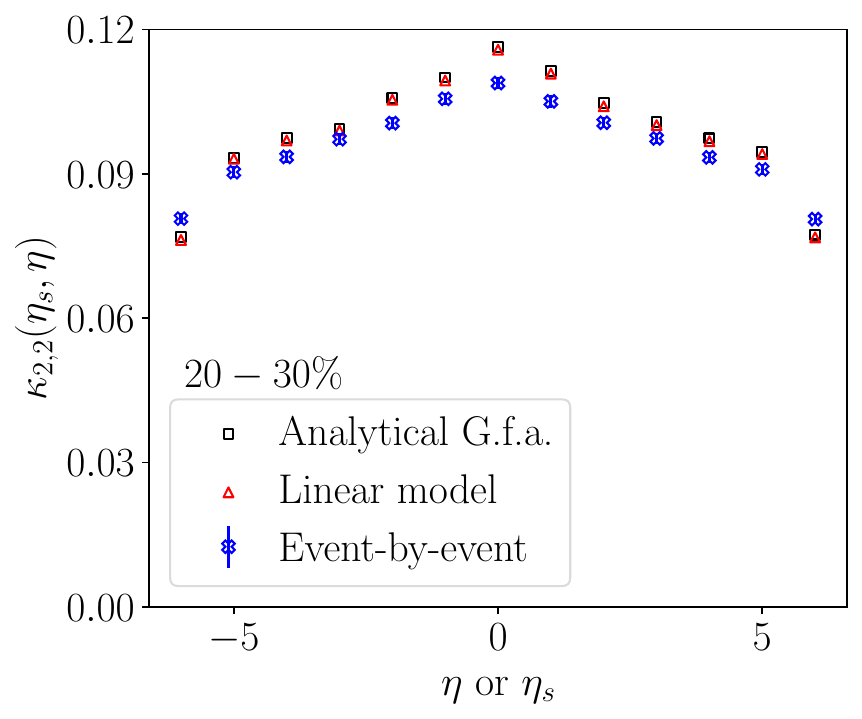}\vspace{-3mm}
	\caption{Linear proportionality coefficient between $v_2(\eta)$ and $\varepsilon_2(\eta_s)$, for $\eta=\eta_s$, as a function of rapidity. The three extraction methods are explained in the text.}
	\label{fig:kappa}
\end{figure}

Using our various approaches to compute observables, it is possible to investigate the rapidity behavior of the linear proportionality coefficient $\kappa_{2,2}(\eta_s,\eta)$ describing the linear (dynamical) response of $v_2(\eta)$ to $\varepsilon_2(\eta_s)$. 
To extract this coefficient, we consider three different approaches in each longitudinal bin, enforcing $\eta_s=\eta$:
\begin{itemize}
 \item For the event-by-event data, we perform a cubic fit as in Eq.~\eqref{eq:kappa} and retain only the linear coefficient.
 
 \item In the linear model, $\kappa_2$ is extracted from a linear fit to the set of $5\times 10^4$ ``linear observables''.

\item In the analytical G.f.a., $\kappa_{2,2}(\eta_s,\eta)$ is obtained from averages using the joint probability distribution:
    \begin{equation*}
        \kappa_{2,2}(\eta_s,\eta)= \frac{\langle v_2(\eta) \rangle}{\langle \varepsilon_2(\eta_s) \rangle} =
        \frac{\displaystyle\int\!v_{2\,} p(\varepsilon_2, v_2) \,\dd\varepsilon_2 \,\dd v_2}
        {\displaystyle\int\!\varepsilon_{2\,} p(\varepsilon_2, v_2)\, \dd\varepsilon_2 \,\dd v_2}.
    \end{equation*}
\end{itemize}
The results are shown in Fig.~\ref{fig:kappa}.
Only the event-by-event data include error bars, estimated using the jackknife method. 
The other approaches involve more intricate uncertainties due to the underlying linear and Gaussian assumptions. 
Again, the results of the linear model and the G.f.a.\ are almost identical, although $\kappa_{2,2}$ is computed with different methods in the two approaches. 
From the values of the sample means in the two rightmost columns of Table~\ref{tab:e2_v2_avg}, one can guess that performing a linear fit to the event-by-event data would yield very similar values to those of the linear approaches.
Here we see that performing the more realistic linear+cubic fit with Eq.~\eqref{eq:kappa} leads to a different value of $\kappa_{2,2}$, which still has the parallel trend as a function of rapidity, namely a decrease by about 25\% when going from midrapidity to $|\eta|=6$.

Instead of using $v_2(\eta)$, it may be more natural to use the value of $v_2$ measured in a longitudinal-momentum rapidity ($y_{\bf p}$), as done in Ref.~\cite{Li:2019eni,Franco:2019ihq} (which study the correlation between $v_2(y_{\bf p})$ and $\varepsilon_2$ from a wider $\eta_s$ range). 
Since $\langle\varepsilon_2(\eta_s)\rangle$ is almost constant across the spacetime-rapidity range used in Fig.~\ref{fig:kappa}, the latter can be seen as representing the (experimentally measurable) pseudorapidity dependence of the average $p_{\rm T}$-integrated $v_2$.

\begin{figure}
	\includegraphics[width=0.9\linewidth]{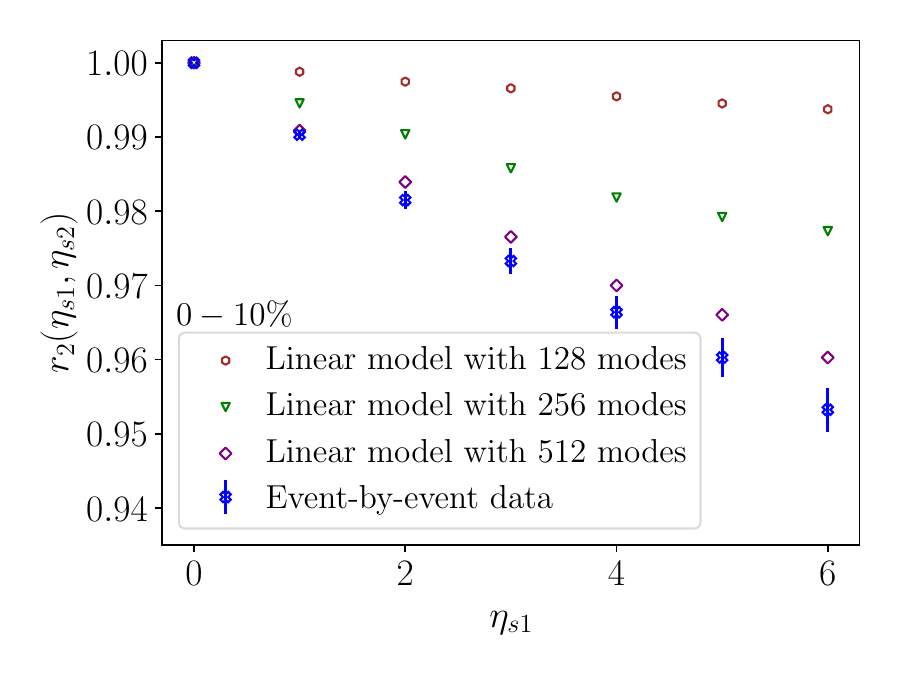}\vspace{-3mm}
	\includegraphics[width=0.9\linewidth]{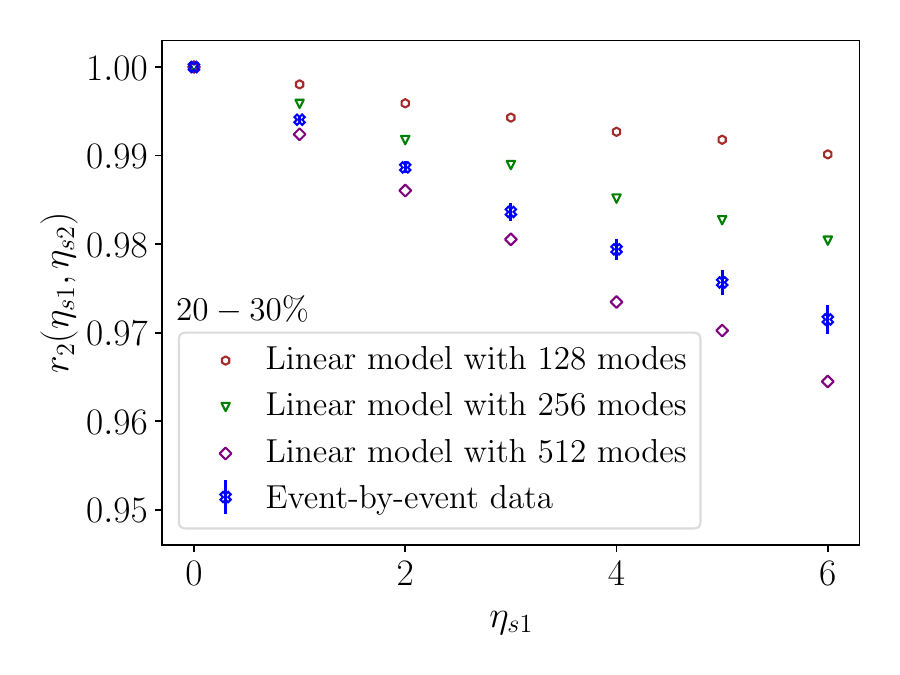}\vspace{-3mm}
	\caption{Longitudinal decorrelation of $\varepsilon_2$, Eq.~\ref{eq:decorrelation}, as function of $\eta_{s1}$ with fixed $\eta_{s2} = 5$, for O+O collisions at 0--10\% (top) and 20--30\% (bottom) centrality. 
	Blue points: event-by-event results from 5000 events. 
	The other markers represent the result within the linear model with $5\times 10^4$ sets of $\varepsilon_2$ values computed using 128, 256, and 512 modes.
	}
	\label{fig:decorrelation}
\end{figure}

Until now, we presented results that can be reproduced with reasonable to very good accuracy using a relatively small set of modes $\{\Psi_l\}$. 
This is however admittedly not the case for all observables. 
In particular, those that crucially depend on a rapidity-asymmetric behavior will naturally involve higher-order modes, like those that are (partly) odd in $\eta_s$ shown in Figs.~\ref{fig:odd_modes_0-10}--\ref{fig:odd_modes_20-30}.
As an example, we display in Fig.~\ref{fig:decorrelation} the longitudinal decorrelation of the initial eccentricity $\varepsilon_2$, defined in analogy to the similar observable for $v_2$~\cite{CMS:2015xmx} as
\begin{equation}
	r_2(\eta_{s1}, \eta_{s2}) = \frac{\langle \varepsilon_{2,\rm c}(-\eta_{s1}) \varepsilon_{2,\rm c}(\eta_{s2}) + \varepsilon_{2,\rm s}(-\eta_{s1}) \varepsilon_{2,\rm s}(\eta_{s2}) \rangle}{\langle \varepsilon_{2,\rm c}(\eta_{s1}) \varepsilon_{2,\rm c}(\eta_{s2}) + \varepsilon_{2,\rm s}(\eta_{s1}) \varepsilon_{2,\rm s}(\eta_{s2}) \rangle},
	\label{eq:decorrelation}
\end{equation}
where we fix $\eta_{s2} = 5$ and vary $\eta_{s1}$.\footnote{We did not look at the longitudinal decorrelation of $v_2$, because it is clear to us that it involves more nonlinear terms than $\varepsilon_2$, due to the dynamical stage: the cubic terms in $\varepsilon_2$ of Eq.~\eqref{eq:kappa}, but possibly also contributions in $\varepsilon_1^2$ or $\varepsilon_1\varepsilon_3$, which involve harmonics that we did not include in this paper.} 
Together with the event-by-event data points (blue, with error bars estimated from the jackknife method), we show successive approximations from the ``linear model'' involving all first 128, 256, or 512 modes. 
Naturally, only those with sizable $\varepsilon_2$ contribute: $\eta_s$-even modes bring $r_2$ closer to 1, while odd ones contribute to the decorrelation.
Looking at the central events (top panel) gives the impression that increasing the number of modes  systematically improves the agreement with event-by-event results. 
However, the bottom panel (20--30\% central events) show that the approach to event-by-event results with the inclusion of more fluctuation modes may not be monotonous, which as stated above should be clear, since modes of opposite parity contribute differently to $r_2(\eta_{s1}, \eta_{s2})$. 
Additionally, since we know from Fig.~\ref{fig:v2_e2_10modes} that the linear model does not give an accurate description of the probability distribution of $\varepsilon_2$, and even less in non-central events, the behavior observed in Fig.~\ref{fig:decorrelation} may be due to this lack of agreement. 
That is, improving the linear model, or a similar approach based on the fluctuation modes, may lead to a better agreement with less modes. 
In any case, it is fair to say that $r_2$ is significantly more demanding than e.g.\ the probability distribution of integrated $v_2$ at midrapidity.

\section{Conclusions}
\label{sec:conclusions}

We have investigated elliptic flow in O+O collisions at $\sqrt{\snn} = 5.36$~TeV within a (non-boost-invariant) 3D \McDipper+\MUSIC\ model, starting from realistic nuclear configurations from NLEFT calculations. 
By decomposing initial states into an average profile and event-by-event fluctuation modes, which is done for the first time for real 3D initial events, we demonstrated that the latter play by far the leading role for elliptic flow and related observables. 
More precisely, we showed that the initial eccentricity $\varepsilon_2$ in O+O collisions is not primarily driven by the departure from azimuthal symmetry caused by a finite impact-parameter value, as in large systems. 
Instead, the main origin of $\varepsilon_2$ is the almost random deposition of (relatively few) hot spots in the transverse plane, caused by the small number of nucleons in the oxygen nuclei, and somewhat independent of the extent of the overlap region (of ``equivalent'' spherical nuclei). 

A first outcome is that the correlation between $\varepsilon_2$ --- and thereby, after the dynamical evolution, $v_2$ --- and $\dd N_{\rm ch}/\dd\eta$ is less marked than in larger systems. 
This is actually visible in the ALICE~\cite{ALICE:2025luc}, ATLAS~\cite{ATLAS:2025nnt} and CMS~\cite{CMS:2025tga} flow results, which show a rather weak dependence on multiplicity in comparison to Pb+Pb or Xe+Xe.
In our study, it means that the fluctuation modes contribute similarly across the two centrality classes we have studied, as exemplified in the trend of the relative weights, or in the joint probability distributions $p(\varepsilon_2,v_2)$.

A second consequence of the importance of initial-state fluctuations is the need for event-by-event simulations in theoretical studies. 
These are computationally expensive, especially in 3D, so that a reduced-cost approach as that used here may prove useful. 
Indeed, the dynamical behavior of the system is encoded in the form of successive response coefficients of observables for chosen modes, from which the probability distribution of the responses in an event sample can then be reconstructed. 
In this paper we restricted ourselves to linear response and we assumed for simplicity that the expansion coefficients $\{c_l\}$ of the fluctuation modes are Gaussian, which is not always true. 
Under these assumptions, we have shown that one can reproduce to good accuracy the bulk of the event-by-event statistical behavior of $v_2$ using only 10 (relevant) fluctuation modes. 
Adding more modes improves the agreement, except for a component which the simplified, linear approach could not hope to capture, namely the nonlinear dependence of $v_2$ at large $\varepsilon_2$, which causes a fatter tail of the elliptic flow distribution towards the larger values. 

Including non-Gaussian $\{c_l\}$ in a numerical approach generalizing the linear-model calculations of this paper is straightforward, while going beyond the linear approximation is also possible: quadratic coefficients for final-state observables are more statistics demanding, which necessitates more oversamplings at particlization~\cite{Krupczak:2025gwg}. 
Yet at that order our fluctuation modes remain independent by construction, so that only one quadratic response coefficient has to be computed for each mode and observable, which remains feasible. 

While some observables can be described with only a few modes, others cannot. 
In particular, here we showed that reproducing the event-by-event value of the longitudinal decorrelation of $\varepsilon_2$ necessitates many more modes, since it is especially sensitive to rapidity-odd modes, which only appear at rather high order in \McDipper.
We did not attempt to optimize the choice of modes that contribute to such observables, which could be done by only considering modes whose relevant response coefficients are larger than some cutoff, but there is room for such improvements in the future. 

Irrespective of this small ``failure'', one should rather appreciate that the expansion in fluctuation modes works and can be successfully exploited. 
This is not obvious, once one realizes that fluctuations are not a ``small perturbation'' in O+O collisions, as we have shown here.

\begin{acknowledgments}
	We would like to thank Gabriel Denicol, Oscar Garc{\'\i}a-Montero, Giuliano Giacalone, Sören Schlichting and Jie Zhu for valuable discussions.
    The authors acknowledge support by the Deutsche Forschungsgemeinschaft (DFG, German Research Foundation) through the CRC-TR 211 'Strong-interaction matter under extreme conditions' - project number 315477589 - TRR 211. 
    H.R is supported by the Research Council of Finland, the Centre of Excellence in Quark Matter, and by the European Research Council (ERC, grant agreements No. ERC-2023-101123801 GlueSatLight and No. ERC-2018-ADG-835105 YoctoLHC).
    The content of this article does not reflect the official opinion of the European Union and responsibility for the information and views expressed therein lies entirely with the authors.
    Numerical simulations presented in this work were performed at the Paderborn Center for Parallel Computing (PC$^2$) and we gratefully acknowledge their support.
\end{acknowledgments}


\end{document}